\documentclass[sigconf]{acmart} %
\usepackage{csquotes}
\makeatletter
\@ifclasswith{acmart}{anonymous}
{%
    \usepackage[true]{anonymous-acm}
}{%
    \usepackage[false]{anonymous-acm} 
}
\makeatother
\usepackage{ifdraft}
\usepackage{etoolbox} 
\usepackage{xspace}
\usepackage{color}
\usepackage{tabularray}
\usepackage{comment}
\usepackage{cleveref}
\renewcommand*{\cref}{\Cref}
\usepackage{hyphenat}
\usepackage{array}
\usepackage{booktabs}
\usepackage{tabularx} 
\usepackage{multirow}
\usepackage{longtable}
\usepackage{subfigure}
\usepackage{soul}
\usepackage{xcolor}
\usepackage{utfsym}
\definecolor{lightyellow}{RGB}{255, 251, 204}

\usepackage{longtable}
\usepackage{colortbl}
\usepackage{multirow}
\usepackage{subcaption}
\usepackage{csquotes}
\usepackage{titlesec}
\titleformat{\section}
  {\normalfont\Large\bfseries\MakeUppercase}{\thesection}{1em}{}

\NewDocumentCommand{\pquote}{+O{} +m}{\blockquote[#1]{\textit{#2}}}

\newcommand{\elide}[1]{\textelp{}} 
 
\newcommand{\textcite}[1]{\citeauthor{#1}~\cite{#1}}

\usepackage{tikz}

\definecolor{lightyellow}{RGB}{255, 251, 204}
\definecolor{lightgreen}{RGB}{204, 255, 204}


\copyrightyear{2024}
\acmYear{2024}
\setcopyright{rightsretained}
\acmConference[CSCW Companion '24]{Companion of the 2024 Computer-Supported Cooperative Work and Social Computing}{November 9--13, 2024}{San Jose, Costa Rica}
\acmBooktitle{Companion of the 2024 Computer-Supported Cooperative Work and Social Computing (CSCW Companion '24), November 9--13, 2024, San Jose, Costa Rica}\acmDOI{10.1145/3678884.3681850}
\acmISBN{979-8-4007-1114-5/24/11}

\begin{document}

\title{Harnessing LLMs for Automated Video Content Analysis: An Exploratory Workflow of Short Videos on Depression}

\author{Jiaying (Lizzy) Liu$^*$}
\email{jiayingliu@utexas.edu}
\orcid{0000-0002-5398-1485}
\affiliation{
  \institution{School of Information, The University of Texas at Austin}
  \country{USA}
 }

\author{Yunlong Wang$^*$}
\email{wang_yunlong@ihpc.a-star.edu.sg}
\orcid{0000-0003-0611-0078}
\affiliation{
  \institution{Institute of High Performance Computing (IHPC), A*STAR}
  \country{Singapore}
 }

\author{Yao Lyu}
\orcid{0000-0003-3962-4868}
\email{yaolyu@psu.edu}
\affiliation{%
  \institution{Pennsylvania State University}
  \city{University Park}
  \state{Pennsylvania}
  \country{USA}
}

\author{Yiheng Su}
\email{sam.su@utexas.edu}
\orcid{0009-0001-8021-429X}
\affiliation{
  \institution{CrowdSourcing Lab, The University of Texas at Austin}
  \country{USA}
 }

\author{Shuo Niu}
\email{shniu@clarku.edu}
\orcid{https://orcid.org/0000-0002-8316-4785}
\affiliation{%
  \institution{Clark University}
  \streetaddress{950 Main St.}
  \city{Worcester}
  \state{MA}
  \country{USA}
  \postcode{01610}
}
\author{Xuhai "Orson" Xu}
\email{xx2489@columbia.edu}
\orcid{0000-0001-5930-3899}
\affiliation{
  \institution{Department of Biomedical Informatics, Columbia University}
  \country{USA}
}

\author{Yan Zhang}
\email{yanz@utexas.edu}
\orcid{0000-0002-1130-0012}
\affiliation{
  \institution{School of Information, The University of Texas at Austin}
  \country{USA}
}

\renewcommand{\shortauthors}{}

\begin{abstract}
Despite the growing interest in leveraging Large Language Models (LLMs) for content analysis, current studies have primarily focused on text-based content. In the present work, we explored the potential of LLMs in assisting video content analysis by conducting a case study that followed a new workflow of LLM-assisted multimodal content analysis. The workflow encompasses codebook design, prompt engineering, LLM processing, and human evaluation. We strategically crafted \textit{annotation prompts} to get LLM \textit{Annotations} in structured form and \textit{explanation prompts} to generate LLM \textit{Explanations} for a better understanding of LLM reasoning and transparency. To test LLM's video annotation capabilities, we analyzed 203 keyframes extracted from 25 YouTube short videos about depression. We compared the LLM Annotations with those of two human coders and found that LLM has higher accuracy in \textit{object} and \textit{activity} Annotations than \textit{emotion} and \textit{genre} Annotations. Moreover, we identified the potential and limitations of LLM's capabilities in annotating videos. Based on the findings, we explore opportunities and challenges for future research and improvements to the workflow. We also discuss ethical concerns surrounding future studies based on LLM-assisted video analysis.
\end{abstract}
\titlespacing*{\section}
  {0pt}{1ex plus 1ex minus .2ex}{0.5ex plus .2ex}


\maketitle
\def\thefootnote{*}\footnotetext{These authors contributed equally to this work}
\section{Introduction}

\begin{figure*}[htbp!] 
    \centering
    \includegraphics[width=.6\linewidth]{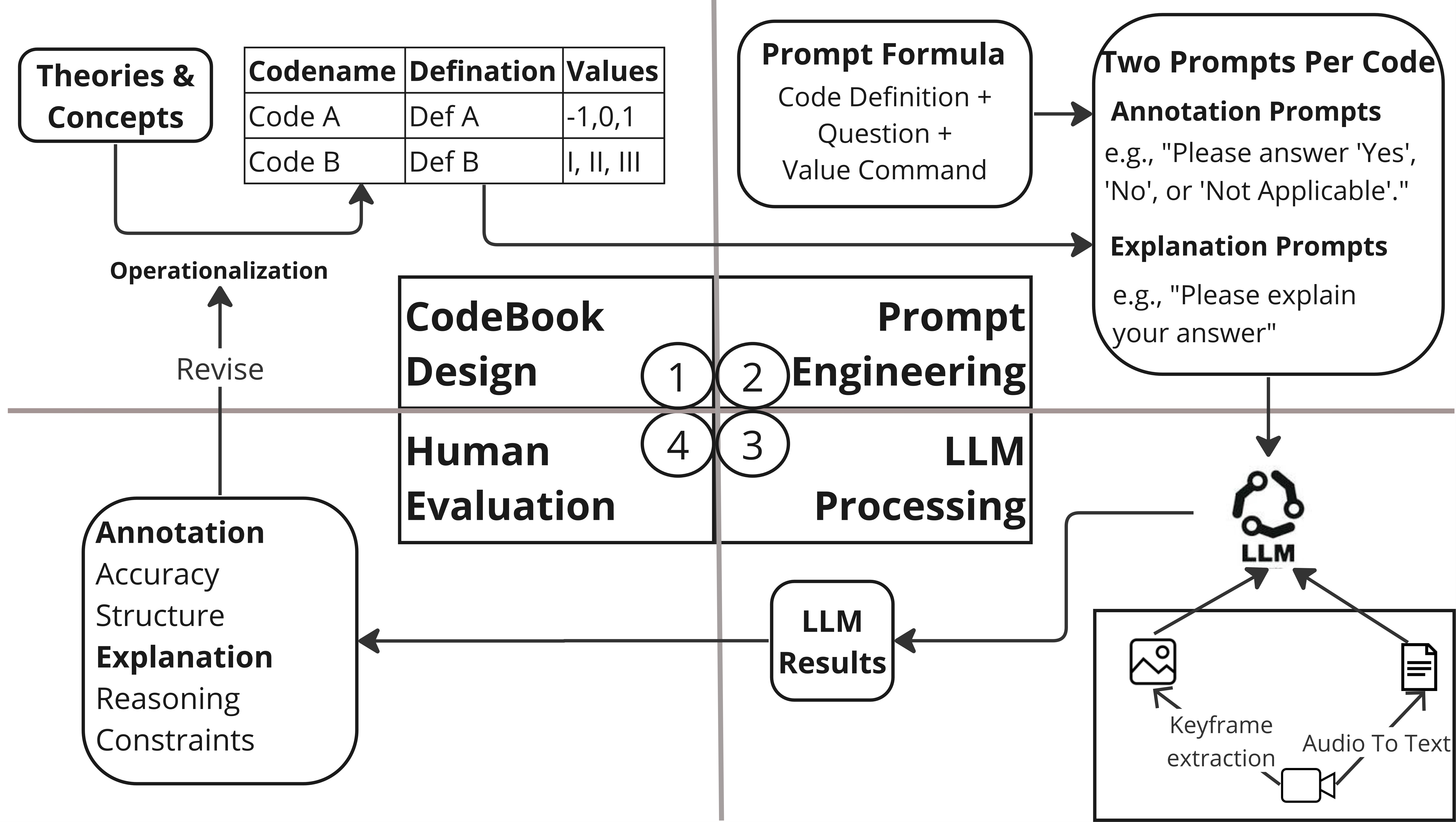}
    \caption{LLMs-Assisted Video Content Analysis Workflow}
    \label{fig:workflow}
\end{figure*}

\citeauthor{}
With the ubiquity of video content across social media, there is a growing need for robust methods to analyze and interpret multimodal data (textual, audio, visual) \cite{bartolome2023literature}. Existing approaches for video analysis include digital ethnography \cite{kubitschko_innovative_2016}, content analysis  \cite{drisko_content_2016}, crowdsourcing \cite{aroyo_crowdsourcing_2019}, and computational methods \cite{biel_vlogsense_2011}. However, each of these methods has certain limitations. Content analysis and digital ethnography are labor-intensive, requiring extensive manual effort from trained researchers, limiting its scalability and efficiency in handling large datasets. Crowdsourcing leverages the collective power of online crowd workers, but it might suffer from the low quality of analysis, possibly due to the lack of worker training \cite{retelny_no_2017}. Computational methods show promise in automating video analysis but still struggle with the complexities of understanding nuanced human behaviors and contextual factors depicted in videos \cite{biel_vlogsense_2011}. 

To address this gap, we propose a LLM-assisted content analysis workflow (Figure \ref{fig:workflow}) that can both save human labor in the analysis and capture the nuances in the content. 
Recently, LLMs have been tested in various types of qualitative data analysis \cite{shen_shaping_2023}, such as deductive thematic analysis \cite{zhang_redefining_2023}; research findings show that LLMs are able to improve coding efficiency \cite{chew_llm-assisted_2023}, facilitate collaborative coding \cite{gao_collabcoder_2023}, and increase inter-coder reliability \cite{xiao_supporting_2023}. However, most of the research is based on text data. In this paper, we join the exploration of LLMs for data analysis and situate it in the video-data context. Using LLMs, we are specifically interested in customizing codebooks and designing prompts to comprehend and capture more fine-grained information to encompass the semantic and contextual information conveyed in the video content.


Our findings contribute to the growing body of research on LLM-supported video content analysis: 1) We introduce a novel workflow that presents the use of LLMs in video content analysis; 2) We foreground the work of customizing data analysis codebook and designing prompts when using LLMs for analysis; 3) We discuss the capabilities as well as limits of the LLMs-assisted video analysis workflow.

\titlespacing*{\section}
  {0pt}{1ex plus 1ex minus .2ex}{0.5ex plus .2ex}
  
\section{Context and Dataset: Depression-\\Related Videos on YouTube}

We plan to use LLMs to analyze the content of YouTube videos related to users' self-disclosure of depression. Online self-disclosure, defined as sharing sensitive personal information in an online public space \cite{misoch_stranger_2015, andalibi_sensitive_2017}, has been extensively studied on text-based platforms such as Twitter \cite{lachmar_mydepressionlookslike_2017}, Weibo \cite{wang_can_2019}, and Reddit \cite{choudhury_mental_2014}. The multimodal nature of video may enable deeper levels of disclosure and facilitate a stronger sense of connection between creators and their audience \cite{liu_modeling_2023, liu2024understandingfacilitatingmentalhealth}. By applying LLMs to extract quantifiable information from the videos automatically, we aim to enable future studies of exploring how people convey information and emotion via self-disclosure videos \cite{chaudoir2010disclosure}. This case study only uses this flow as a preliminary method to analyze video content to explore the capability and limitations of LLMs in content analysis.


Using the search query "depression" with the YouTube Data API, we collected the metadata (e.g., title, channel, duration) of 3,892 videos uploaded in February 2024.  
We randomly selected 25 videos and extracted 203 keyframes from them for this case study.

\section{LLM-Assisted Video Annotation Workflow and Study Process}
We designed an LLM-assisted workflow to facilitate the content analysis process, together with a human-in-the-loop mechanism to ensure its validity. The workflow contains four steps, as shown in Figure \ref{fig:workflow}.

\subsection{Step 1: Codebook Design}
This step requires researchers to operationalize theories and concepts. This is a crucial step in transforming vague themes into identifiable codes that can be observed in videos. Therefore, we need to seductively generate a codebook that LLMs can use to annotate video information. To ensure the creation of clear and effective prompts for LLMs annotation, the codebook explicitly defines each code's name, definition, and possible values.

In our study, we followed the disclosure processes model \cite{chaudoir2010disclosure} to develop a codebook that captures several key aspects of the concept. Depth of self-disclosure pertains to the extent to which disclosed information is considered highly private or intimate \cite{altman_social_1973}. For example, disclosing in the presence of other people may suggest deeper disclosure, and crying may indicate more intimate disclosure than talking. The codebook also includes the emotional valence of self-disclosure (e.g., positive, negative) \cite{reis_intimacy_1988} and the communication style of self-disclosure (e.g., presenting, interacting) \cite{niu_stayhome_2021}. Table \ref{tab:evaluation} shows the example codes employed in this case study, demonstrating the application of our codebook design process and its role in guiding the development of prompts that capture the nuanced aspects of depth of self-disclosure in depression-related YouTube videos.

\subsection{Step 2: Prompt Engineering}
This step aims to design effective prompts that solicit two types of responses for each code from LLMs: structured Annotations and free-text Explanations, enabling researchers to conduct quantitative analyses while also gaining insights into the LLMs' reasoning processes.
We went through an iterative process of prompt engineering to modify the prompts to get the intended responses. We summarized the strategies we used to design the two types of prompts below.

\begin{itemize}
\item Annotation Prompt: "\textbf{Annotation Prompts}" are closed-ended questions that require LLM to respond with \textbf{Annotations}, which are structured answers for future quantitative analysis, such as predefined multiple choices or Yes/No responses. Each annotation prompt is formulated as \textbf{Code Definition} + \textbf{Question} + \textbf{Value Command}. The Code Definition clearly delineates the scope of the intended responses, ensuring that the LLM focuses on the specific aspect of depth of self-disclosure being addressed. The Annotation Question presents a specific inquiry related to the code, guiding the LLM to provide a targeted response. The Value Command specifies the desired values and formats of the responses, facilitating the extraction of structured output for the Annotation Questions. For example, to annotate whether there is talking behavior, the Annotation Prompt can be, "\textit{Talking behavior refers to the act of verbal communication through spoken language. [Code Definition] + Is there talking behavior in the picture?}" [Question] + "\textit{Please only respond 'Yes', 'No', or 'Not Applicable'.}"[Value Command]. See Table \ref{tab:evaluation} for more examples.
\item Explanation Prompt: "\textbf{Explanation Prompts}" require LLM to respond with \textbf{Explanations} that elaborate the rationales of Annotations and the decision-making process, providing explainability and facilitating human validation. Explanation Prompts are used together with Annotation Prompts to obtain explainable results with structured outputs. Taking the example in the Annotation Prompt section, the corresponding Explanation Prompt would be "\textit{Talking behavior refers to the act of verbal communication through spoken language. [Code Definition] + Is there talking behavior in the picture?" [Question] + Please answer this question with an explanation.}" 
\end{itemize}


\begin{table*}[h]
\caption{Codebook Design, Prompt Engineering, and Human Evaluation}
\label{tab:evaluation}
\small
\begin{tblr}{
width = \textwidth,
colspec = {|Q[l,0.07]|Q[l,0.09]|Q[l,0.4]|Q[l,0.4]|Q[l,0.2]|Q[r,0.08]|Q[r,0.08]|},
hlines,
row{1} = {font=\bfseries,c,gray9},
}
Code Type& Code Name & Code Definition & Question & Value Command & ICR-Two Coders & ICR-LLMs vs Ground Truth \\
\SetCell[r=2]{m}{Object} & N. People & Number of people. & How many people are in the picture? & Arabic numerals. & \textbf{88.35\%} &\textbf{79.80\%} \\
& Food & The presence of food or beverage.  & Is there food or beverages in the picture? & \SetCell[r=6]{m}{'Yes', 'No' or 'Not Applicable'.}  & \textbf{95.63\%} & \textbf{92.12\%} \\
\SetCell[r=3]{m}{Behavior}& Talking & The act of verbal communication through spoken language.  & Is there talking behavior in the picture? & & \textbf{86.41\%} & \textbf{75.37\%} \\
& Crying & The act of shedding tears. & Is there crying behavior in the picture? & & \textbf{98.54\%} & \textbf{93.10\%} \\
& Treatment & The act of providing medical care or health intervention. & Is there treatment behavior in the picture? & & \textbf{96.12\%} & 73.89\% \\
\SetCell[r=2]{m}{Genre} & Presenting & The delivery of information, typically accompanied by visual aids like slides or graphics. It's commonly employed for educational purposes, business presentations, lectures, or seminars. & Does this picture communicate in a presenting style? &  & 70.87\% & 62.07\% \\
& Interacting & Interacting style refers to creators establishing a simulated interpersonal relationship with their audience, fostering a sense of engagement and connection. & Does this picture communicate in an interacting style? & & 60.68\% & 57.14\% \\
\SetCell[r=1]{m}{Emotion} & Valence & The positive or negative character of an emotion.  & What is the emotional valence of the picture? & 'Positive', 'Negative', 'Hard to distinguish' or 'Not Applicable'. & \textbf{79.13\%} & 32.51\% \\
\end{tblr}
\raggedright \textbf{Note:} ICR (Intercoder Reliability) is a measure of the agreement between two or more coders when annotating the same data independently. We calculate ICR using the Percentage method \cite{o2020intercoder}, which determines the percentage of entries where one coding decision is in agreement with another. We \textbf{bold} the ICRs higher than 75\% as acceptable \cite{burla2008text}. 
\end{table*}

\subsection{Step 3: LLMs Processing}
This step unfolds the process of decomposing videos into keyframes and transcripts and using LLMs for keyframe and transcript annotation as illustrated in Figure \ref{fig:workflow}.
\begin{itemize}
\item Deciding Types of LLM: Video LLM, such as PandaGpt and V-Llama, have shown promise in directly comprehending video inputs. However, recent studies indicated that the performance of these Video LLMs was inferior to state-of-the-art Image LLM \cite{liu_tempcompass_2024}. We tested 5 videos on V-Llama and found that it struggles to capture the nuances of video content, making it difficult to validate (e.g., identifying specific activities at particular timestamps). In contrast, Image LLM demonstrated better performance and robustness in image comprehension. Text LLMs have also shown maturity, and prior studies suggested promising results in fulfilling qualitative analysis tasks.
\item Decomposing Videos into Keyframes and Transcripts: Consequently, our workflow decomposes videos into images, audio, and transcripts, incorporating multiple LLM to annotate various content modalities. We applied ffmpeg \cite{tomar2006converting} to extract keyframes (i.e., independently coded frames) from the video data. This approach allows us to leverage the strengths of Image and Text LLM for comprehensive video content analysis.
\item Selection of LLMs: After trying several models such as BLIP-2 \cite{li2023blip2} and LLaVA-1.5 \cite{liu2024improved}, we selected "llava-v1.6-mistral-7b-hf" \cite{liu_improved_2023} (one of the SOTA Image LLMs) to analyze keyframes. We will use LLM to refer to this LLaVa-1.6 model hereafter when reporting the results. We used the default parameter settings in the model. Even though this model has demonstrated superior performance for image understanding \cite{zhang2024llavar, lin2023videollava}, it is unknown how it would perform in our task. The present workflow integrates other models to extract information from video transcripts and audio. By combining the annotations generated from both keyframes and transcripts, our workflow enables a more comprehensive understanding of the video content. The keyframes provide visual context, capturing the setting, objects, and actions depicted in the video, while the transcripts and audio offer insights into the spoken words and themes conveyed through language. However, due to the limited space of this paper, we only focus on the image comprehension part.
\end{itemize}

\subsection{Step 4: Human Evaluation}
To evaluate LLM results, we employ an iterative process of human evaluation and adjust the codebook and prompts accordingly. This human-in-the-loop process ensures that the final annotations are reliable and well-aligned with human judgments, enhancing the validity of the research findings derived from the annotated data.

\begin{itemize}
\item Human coding. In our case study, two human coders independently coded the data and reconciled their decisions to generate a "ground truth" decision, which serves as the benchmark for evaluating the LLM's responses. 
\item Quantitative Evaluation. We then calculated the percentage of agreement between the LLM's responses and the ground truth. 
\item Qualitative Evaluation. We examined the Explanations of Annotations and qualitatively analyzed whether the LLM provides logical reasoning for its decisions. This step could help identify potential areas for improvement in the codebook and prompts by assessing the coherence and validity of the model's reasoning process.
\end{itemize}


\section{Study Results} 

We report our findings on LLM's potential (marked with \usym{1F5F8}) and limitations (marked with {\usym{2753}}) in video content analysis by evaluating LLM generated Annotations and Explanations. We note that the analysis below is based on the results using the LLaVA-1.6 model as we introduced in the previous section.

\subsection{Examination of LLM Annotation}
We selected examples from four types of codes that we used LLM to annotate (i.e., \textbf{object annotation, behavior annotation, emotion annotation, and concept annotation}). Object annotation requires the LLM to identify the presence or number of entities within the image, such as humans and food. Behavior annotation involves identifying specific activities or actions the individuals perform in the video. Genre and Emotion annotation needs a deeper understanding of the communication styles, the underlying context, and emotional expressions represented in the video. We compared LLM Annotations with human annotations. As shown in Table \ref{tab:evaluation}, we report the Intercoder Reliability (ICR) \cite{o2020intercoder} between two human coders, as well as between LLM and human-labeled Ground Truth, which is the reconciled human decision, on the four types of questions. 

\subsubsection{Structured Output} Our prompting methods can elicit LLM to generate structured annotations of videos to assist quantitative data analysis. 

\usym{1F5F8} \textbf{LLM can generate structured output when using effective prompts.} After iterative prompt crafting, we found that adding quotation marks to the potential values can improve the LLM's adherence to the specified format. For example, prompts like "\textit{Answer positive, negative, hard to distinguish, or not applicable}" may render free-text responses, which was difficult for further quantitative analysis. After redesigning the Annotation Prompts by adding the quotation, e.g., using  "\textit{Answer 'Positive', 'Negative', 'Hard to distinguish', or 'Not Applicable'}", the responses become clean and structured.

\begin{figure*}[htbp!]
  \centering
  \subfigure[An example of Connecting Visuals and Embedded Texts]{\includegraphics[width=0.15\linewidth]{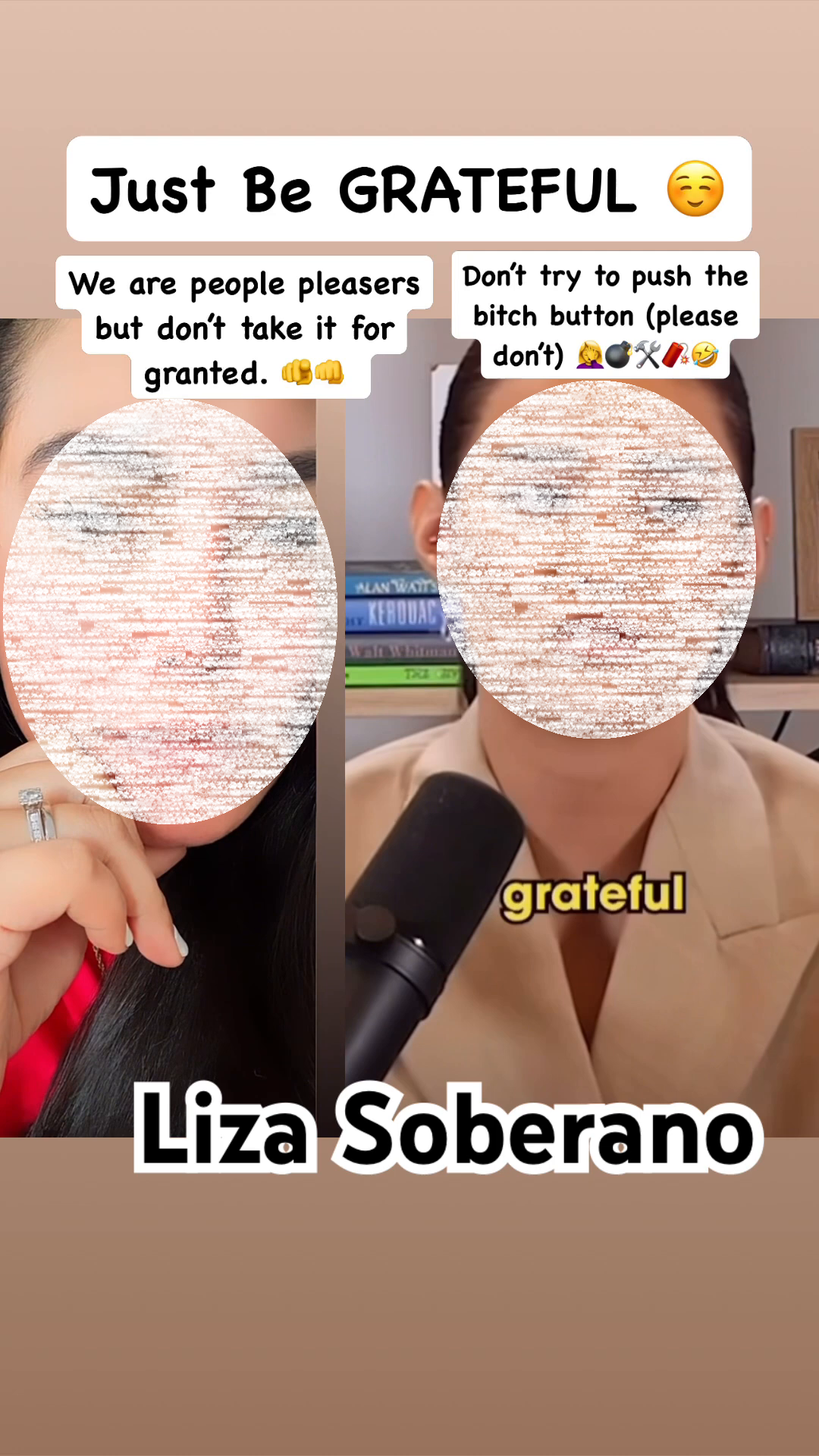}\label{fig:example1}} 
  \hspace{.7em}
  \subfigure[An example of Unfolding Context and Explaining Rationales and Acknowledging Limitations of Annotation]{\includegraphics[width=0.15\linewidth]{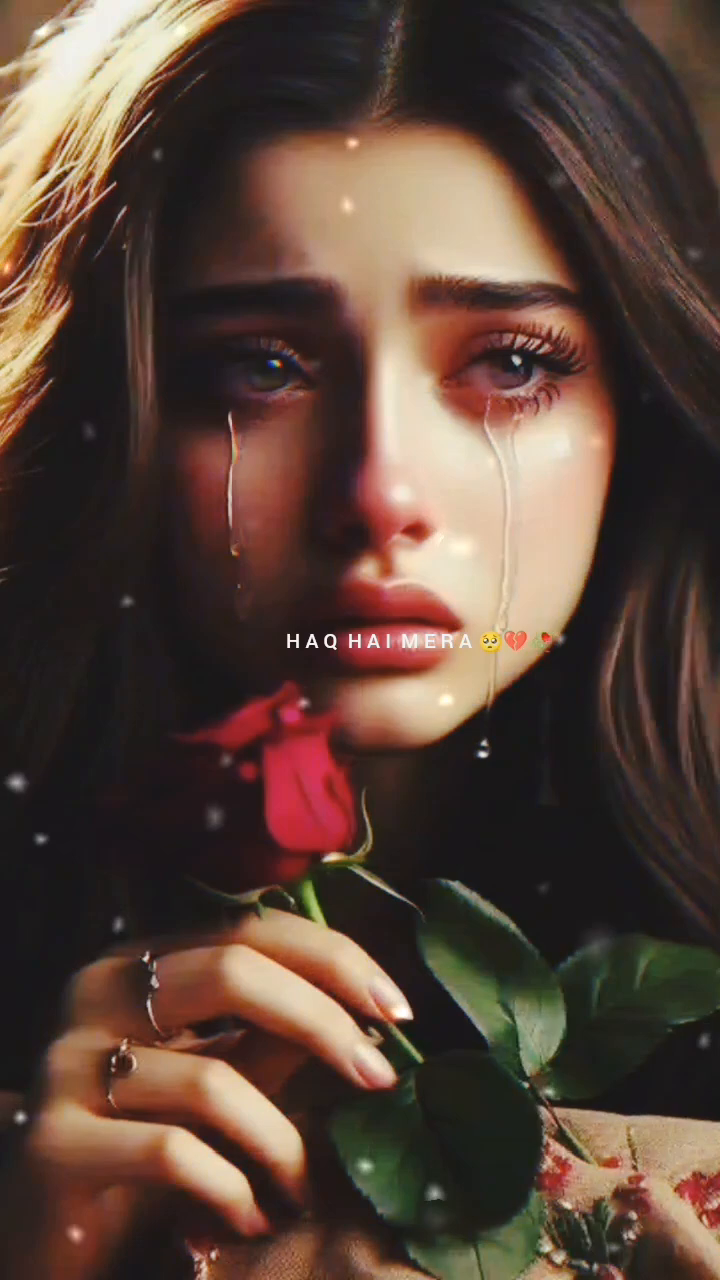}\label{fig:example3}}
  \hspace{.7em}
  \subfigure[An example of Recognizing "Not Applicable" for the "Emotion" code.]{\includegraphics[width=0.15\linewidth]{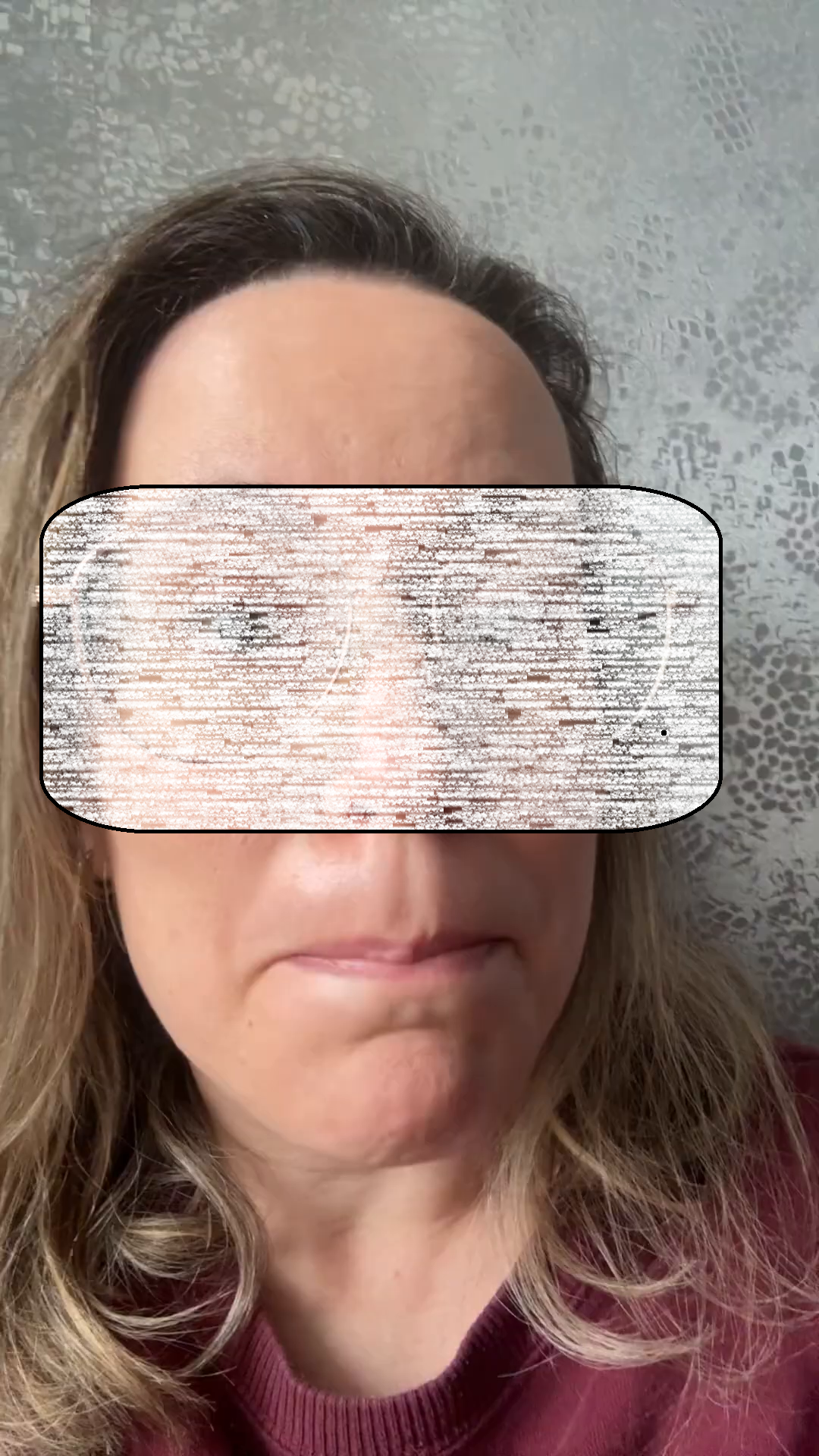} \label{fig:example4}}
  \hspace{.7em}
  \subfigure[An example of Assisting Comprehension of Non-English Content]{\includegraphics[width=0.15\linewidth]{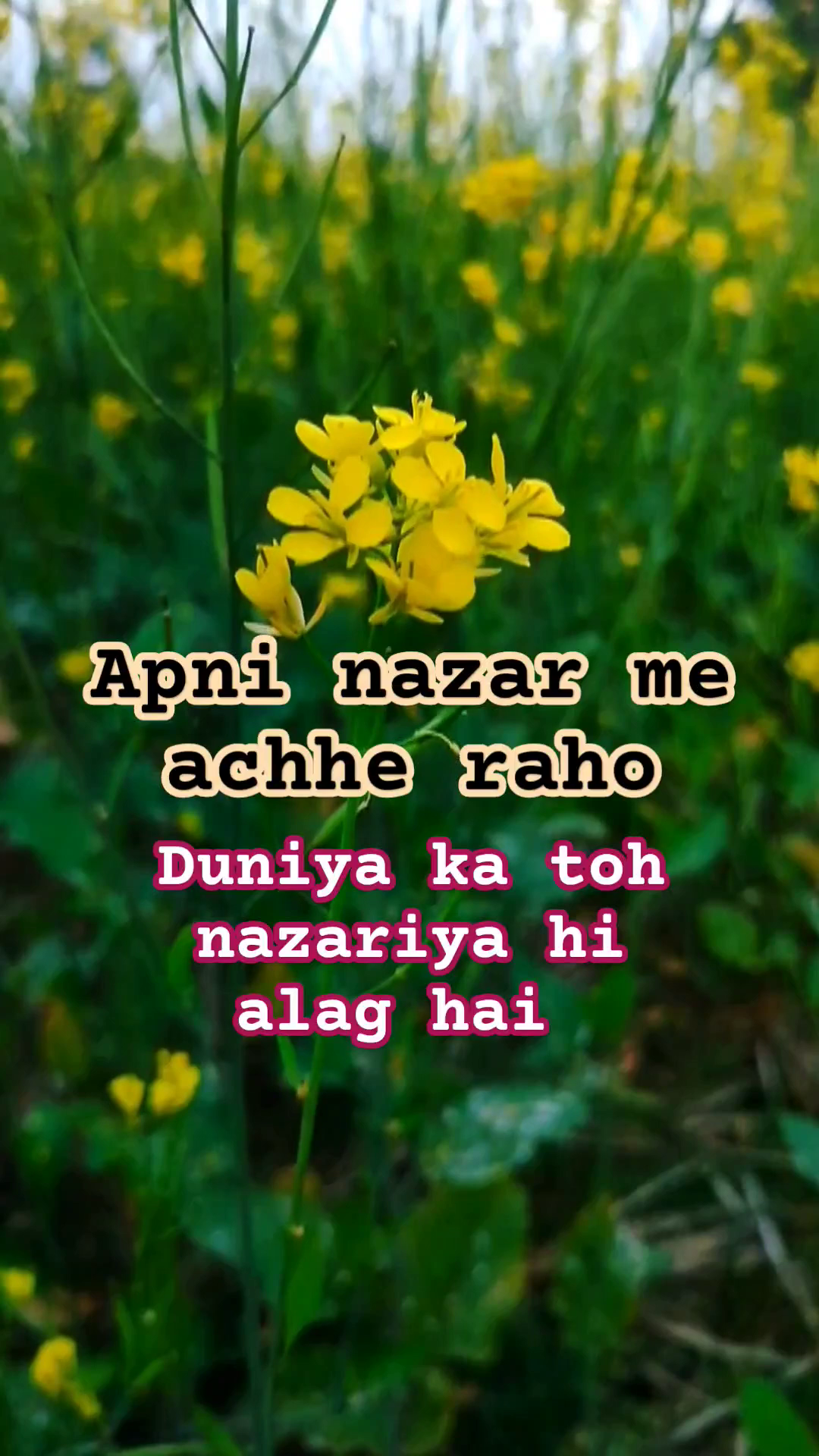}\label{fig:example2}} 
  \caption{Examples of Keyframes}
\end{figure*}
\subsubsection{Accuracy of Annotations} LLM shows various capabilities in annotating different types of codes.

\usym{1F5F8} \textbf{LLM is able to annotate Objects and Behaviors with acceptable accuracy.} When comparing the ICR between the two human coders and the LLM, we find that the LLM achieves higher reliability scores for \textit{object} and \textit{activity} Annotations. Except for the treatment behavior, the ICRs of LLM vs Ground Truth are all above 75\%. However, the LLM's performance is relatively lower for \textit{genre} Annotations (e.g., Presenting genre: 62.07\% and Interacting genre: 57.14\%) and \textit{emotion} (32.51\%). These results suggest that LLMs may be more reliable when annotating concrete and observable elements in keyframes, such as objects and activities, but struggle with abstract and subjective concepts like emotions and genres.


{\usym{2753}} \textbf{LLM has Limited Ability in Understanding Emotions and Genres.} In our initial test, the accuracy of Emotion and Concept Annotation was lower compared to Object and Behavior Annotation. LLM may struggle with accurately discerning emotional states and extracting abstract concepts, which require a more nuanced understanding of the video's content and context. It is worth noting that the ICR between the two human coders is also lower for \textit{genre} Annotations (e.g., Presenting genre: 70.87\% and Interacting genre: 60.68\%) and \textit{emotion} (79.13\%). This highlights the inherent difficulty in achieving high agreement for these more complex and interpretive annotation tasks, even among human coders. In our initial test, the accuracy of Emotion and Concept Annotation was lower compared to Object and Behavior Annotation. LLM may struggle with accurately discerning emotional states and extracting abstract concepts, which require a more nuanced and holistic understanding of the video's content and context. 

\subsection{Examination of LLM Explanation}
We qualitatively examined the Explanations of LLM's annotation rationales (i.e., responses to Explanation Prompts) and revealed some challenges and opportunities in LLM's content analysis.

\subsubsection{Reasoning in Explanation} To assess how the LLM annotated the codes, we examine the reasoning of LLM Explanations and report its abilities in making sense of human-defined codes and connecting with keyframes. 

\usym{1F5F8} \textbf{LLM showed potential in appropriately understanding Code Definitions}: LLM show potential in understanding human-described code definitions. Codes about objects and human behaviors can be comprehended without specific definitions (e.g., food and crying). For more abstract codes, LLM can capture the nuances, as shown in the explanation for the "Presenting" code: "\textit{A presenting style in a traditional sense would involve bullet points, text, and possibly images that are relevant to the topic being discussed. The image you've shown is more of a visual narrative or a piece of art that could be used as a visual aid in a presentation, but it does not follow the standard layout or structure of a presentation slide.}"

\usym{1F5F8} \textbf{LLM could unfold image context and explain rationales}: LLM can provide reasonable explanations by analyzing the relationships between keyframe content and code definitions. For example, the "Presenting" code of Figure \ref{fig:example3} is annotated as "No" because "\textit{The image you've provided appears to be a digital artwork or a stylized photograph rather than a slide from a presentation. It features a close-up of a person's face with a rose and a tear, which is not a typical format for a presentation slide.}" In another case, when explaining why it responds with "Yes" for the "Interacting" code in Figure \ref{fig:example1}, it states, "\textit{The image also includes a person's face, which is a common way for creators to engage with their audience by showing their own face and possibly their emotions or reactions. This can help to build a personal connection with the viewers.}" The explanations typically start with a summary sentence of the image content before delving into the details that justify the annotations, providing both factual and argumentative information.

\usym{1F5F8} \textbf{LLM could connect visuals and embedded texts}: Many short videos in our dataset use embedded text alongside visuals, which is also the style of other topics of short videos. In these cases, it is crucial to capture the embedded text to make sense of the images, as relying solely on the transcripts may not provide sufficient context. LLM can automatically extract the embedded text and relate the visuals and texts. For example, in Figure \ref{fig:example1}, "\textit{The text overlay on the image includes a message that says, 'Just Be GRATEFUL' and a series of emojis that seem to be part of a conversation or a challenge. The text is written in a casual, conversational tone, which is typical of an interacting style.}"

\usym{1F5F8} \textbf{LLM could assist comprehension of Non-English content}: Although we did not include prompts to request the LLM to translate text overlays, the LLM Explanations often contained not only the translation to English but also the interpretation of the quote in its corresponding cultural context. For example, in Figure \ref{fig:example2}, it elaborated, "\textit{the text overlay reads 'Apni nazar me ache raho Duniyaya ka toh nazariya hi alag hai,' which is a Hindi phrase that translates to 'Look at your own world, the world is different from what you see. This phrase is often used as a metaphor for focusing on the present moment.}'" This demonstrates the potential of LLMs to handle multi-lingual content and provide culturally and contextually relevant explanations.

\usym{2753} \textbf{LLM may mix reasoning in long Explanations}: The current Explanations typically describe the keyframes before elaborating on the reasoning. These descriptions can be long and detailed, hindering readability when assessing the logic of the Explanations. For example, when explaining its Annotation of emotion floor in Figure \ref{fig:example2}, the LLM starts with, "\textit{The image shows a close-up of a flower, specifically a yellow flower, with a blurred background of green foliage. The flowers are in full bloom and are the main focus of the image.}" This description does not entail direct reasoning until the next paragraph, making it difficult for human evaluators to quickly grasp the LLM's thought process.

\usym{2753} \textbf{LLM sometimes showed hallucination and gave conflicting Annotation and Explanation}: In some cases, there are discrepancies between the LLM's Annotation and Explanation. For example, in the keyframe presented in Figure \ref{fig:example1}, the Annotation is "\textit{No}". However, the Explanation reads, "\textit{Yes, the person in the image is directly talking to the audience. They are likely addressing a group of people or giving a presentation. The presence of a microphone suggests that they are speaking in a public setting, such as a conference or a meeting, where amplification is necessary to ensure that everyone can hear them clearly.}" This inconsistency highlights the need for further refinement of the LLM's reasoning capabilities and the importance of human evaluation in identifying such issues.

\subsubsection{Awareness of Constraints} The LLM can recognize instances where the annotation prompts may not be directly relevant or applicable to the image content. This capability is crucial for maintaining the integrity of the analysis process and avoiding forced or irrelevant annotations.

\usym{1F5F8} \textbf{LLM showed potential in recognizing "Not Applicable" cases}: LLM can identify instances where the annotation prompts may not be directly applicable to the image content. For example, Figure \ref{fig:example4} is annotated as "Not Applicable" for the code "Valence" of emotion. It explains, "\textit{The person in the picture appears to be displaying a slightly puzzled expression. The eyebrows are slightly furrowed, and the mouth is closed with a slight pucker, which could indicate a moment of thoughtfulness or mild confusion. It is not possible to definitively determine the emotions involved in the picture without additional context.}"

\usym{1F5F8} \textbf{LLM showed potential in acknowledging limitations of Annotation}: Many explanations suggest LLM's awareness of the potential limitations or errors in annotations. For example, the LLM explanation of Figure \ref{fig:example3} states, "\textit{The content of the image shows a character with tears and a distressed expression, which \textbf{could be} interpreted as a depiction of emotional distress or suffering. \textbf{However, it is important to note that the context of the scene and the character's intentions are not clear from the image alone.}}" It often attached a sentence declaring its Explanations to Annotations of emotions, "\textit{Please note that the emotions listed are based on the content of the text and the context of the situation described.}" This demonstrates the LLM's ability to acknowledge the limitations of the information available in a single image.

\section{Discussion and Future Work}
The contribution of this work is two-fold: a new workflow of LLM-assisted video content analysis, and an empirical study to examine the LLM's ability to annotate video content. We note the limitations of this work, including the small sample of short videos on the topic of depression and the limited number of LLMs being applied. Despite these limitations, our study still sheds light on how to apply LLMs for video content analysis. We defer more comprehensive evaluation to future work. 

\subsection{Future Opportunities and Challenges}

LLMs could reduce the human labor required for manual coding in video content analysis. Researchers can create a structured framework (e.g., a codebook) to guide LLMs to automatically identify and extract meaningful insights from unstructured video data such as User Created Content (UGC) in video-sharing platforms and lab/field recordings and eye-tracking data in contexts such as Human-Robot Interactions and Virtual Reality \cite{hauser2024vid2realhrialignvideobased, zhang_design_2023}. In our case, operationalizing the concept of depth of self-disclosure into various observable behaviors in different settings, such as the presence of other people and the behaviors disclosed, allows us to understand the nuances of how individuals express themselves in video content. However, we find the most critical step in this process is operationalizing concepts into well-defined codes, as it sets the foundation for accurate and comprehensive analysis by LLMs. Further, it is difficult for researchers to curate prompts to harness the power of LLMs (e.g., annotating abstract concepts such as genre in images, as shown in our case study) and effectively evaluate the structured output generated by LLMs. Future work can investigate how to better involve humans in the workflow. 

\subsection{Directions for Workflow Improvement}
Currently, using single keyframes for analysis is limited in understanding the dynamic context in videos. This might lead to the low Inter-Coder Reliability scores for emotion and genre between the two coders. Emotions and genres often require a more comprehensive understanding of the video's context and narrative, which may not be reflected in isolated keyframes by human coders or LLMs. Future work could explore the ability of improved LLMs that directly comprehend videos or methods of combining keyframes to provide more contextual information. 

Combining analysis from multiple information modalities is a promising avenue for future research. In our current workflow, we decode videos into transcripts and keyframes, but integrating the results from these multiple sources remains challenging. For example, the code of "emotion" may be indicated through both keyframes and transcripts. Feeding the surrounding transcripts together with the images could enable better comprehension of the visual information. Additionally, understanding the temporal features of videos is crucial. By matching the transcripts and keyframes according to timestamps, our next step is to explore methods to capture the continuity of keyframes. 
\subsection{Ethical Considerations}
While video data becomes increasingly accessible for research purposes, this study raises ethical concerns about influencing people's intentions to post videos in public spaces. Individuals who share personal experiences or sensitive information through videos may feel that their content is being exploited \cite{pendse2024advancing}. It is crucial to develop ethical guidelines and protocols to ensure that video data is collected, analyzed, and used responsibly, with proper measures in place to protect the privacy and dignity of the individuals involved. Researchers must also be mindful of the potential impact of their work on the communities they study.
\section{Acknowledgement}
We want to acknowledge Dr. Ciaran B. Trace, Shelby Devin, Olivia Frolichstein, Anabiya Momin, and Yaning Zhu's early contribution to the codebook development of this project.

\bibliographystyle{ACM-Reference-Format}
\bibliography{jiaying-ref,video,LLM, otherbibs}
\appendix

\end{document}